\documentclass[reprint,amsmath,amssymb,aps, onecolumn, prl]{revtex4-2}

\usepackage{amsmath,amssymb}
\usepackage{geometry}
\usepackage{graphicx}
\usepackage{bm}
\usepackage{wasysym}
\usepackage{bm}
\usepackage{subcaption}

\begin{document}



\title{A Superconducting Levitating Oscillator with $<1\,\mu$Hz Resonance Linewidth}

\author{M. Array\'{a}s}
\affiliation{\'{A}rea de Electromagnetismo, Universidad Rey Juan Carlos, Tulip\'{a}n s/n, 28933, M\'{o}stoles, Madrid, Spain}
\author{J. Clothier}
\affiliation{Department of Physics, Lancaster University, LA1 4YB, United Kingdom}
\author{C. C. E. Elmy}
\affiliation{Department of Physics, Lancaster University, LA1 4YB, United Kingdom}

\author{\v{S}. Midlik}
\affiliation{Faculty of Mathematics and Physics, Charles University, Ke Karlovu 3, 121 16, Prague, Czech Republic}
\affiliation{Department of Physics, Lancaster University, LA1 4YB, United Kingdom}
\author{R. Schanen}
\affiliation{Department of Physics, Lancaster University, LA1 4YB, United Kingdom}
\author{J. L. Trueba}
\affiliation{\'{A}rea de Electromagnetismo, Universidad Rey Juan Carlos, Tulip\'{a}n s/n, 28933, M\'{o}stoles, Madrid, Spain}
\author{C. Uriarte}
\affiliation{\'{A}rea de Electromagnetismo, Universidad Rey Juan Carlos, Tulip\'{a}n s/n, 28933, M\'{o}stoles, Madrid, Spain}
\author{D. E. Zmeev}
\affiliation{Department of Physics, Lancaster University, LA1 4YB, United Kingdom}

\date{\today}

\begin{abstract}
  Experiments aimed at quantifying the interface between quantum and classical physics necessarily require a high degree of isolation from the environment: wavefunction collapse and quantum gravity effects at laboratory scales are predicted to be very subtle. Ideally, such tests would be performed in a closed system at extremely low temperatures in order to rule out any external influence and thermal fluctuations. Cryogenic levitated macroscopic bodies are excellent candidates for an accurate laboratory approximation of such systems, as a tether to another body would violate the requirement for the system to be fully closed.  Here we report a significant milestone on the way to a practically suitable approximation of such closed system. We have built a milligram-mass superconducting oscillator operating at millikelvin temperatures showing extremely low dissipation rate, with the oscillator ring-down time exceeding 110 hours. This corresponds to the resonance linewidth of less than 0.8\,$\mu$Hz. The experimental setup is highly tunable and is compatible with adiabatic nuclear demagnetisation, promising even lower temperatures and lower dissipation levels in the future. We demonstrate the capability of our device by measuring drag from $^3$He impurities in superfluid $^4$He at a level of $\sim10^{-8}$ with the drag force in the femtonewton range.
\end{abstract}

\maketitle

\section{Introduction}

Levitating superconductors have been at the forefront of precise measurements of small effects. Some impressive examples include measurements of minute changes of gravity on Earth measured using displacements of a stationary superconducting sphere~\cite{goodkind1999superconducting}, and measurements of the effects of general relativity in space using gyroscopes comprised of spinning superconducting spheres~\cite{will2015focus}.  Oscillators based on the effect of superconducting levitation demonstrate a great promise in achieving a high precision of measurements as well~\cite{gonzalez2021levitodynamics, cirio2012quantum, Hofer_Gross_Higgins_Huebl_Kieler_Kleiner_Koelle_Schmidt_Slater_Trupke_Uhl_Weimann_Wieczorek_Aspelmeyer_2023, fuchs2024measuring}. Some uses of precise measurements using levitating oscillators include magnetometry~\cite{ahrens2025levitated} and accelerometry~\cite{timberlake2019acceleration}. A  macroscopic mechanical oscillator prepared in a quantum state would be instrumental in the experiments attempting to resolve the question whether gravity is quantum~\cite{carney2019tabletop, lami2024testing}.  Such oscillator would have to be sensitive to extremely small forces.


Achieving the longest possible coherence time $\tau~$ in force sensors based on mechanical oscillators is crucial for improving precision of measurements as the noise spectral density is given by~\cite{vinante2020ultralow}:
\begin{equation}
    S_F = \frac{8k_BMT}{\tau},
\end{equation}
with $T$ being the temperature of the surrounding bath and $M$ the mass of the oscillator.
Therefore, a figure of merit used for ultimate noise characterization of such force sensor is the ratio $T/\tau$. In the absence of phase decoherence sources (e.g. time-variable levitation potential), the coherence time is equal to the ring-down time. Perhaps the longest $\tau\sim2\times 10^4$\,s observed for any superconducting oscillator before this work was for a tin-lead alloy ball magnetically levitated in a dilution refrigerator at 15\,mK~\cite{Hofer_Gross_Higgins_Huebl_Kieler_Kleiner_Koelle_Schmidt_Slater_Trupke_Uhl_Weimann_Wieczorek_Aspelmeyer_2023} with $T/\tau\sim 10^{-6}$\,K\,s$^{-1}$. Longer times of $\tau=10^5$\,s have been observed for a ferromagnet levitated in a lead trap \cite{fuchs2024measuring}. The bath temperature in the latter system is hard to measure as lead becomes a poor thermal conductor below its transition at 7.2\,K, resulting in an estimated $T/\tau\sim 10^{-5}$\,K\,s$^{-1}$. Here we present an oscillator with $T/\tau\sim 10^{-8}$\,K\,s$^{-1}$, with a direct path to take our sensor to the sub-millikelvin regime and achieving $T/\tau\sim10^{-10}$\,K\,s$^{-1}$, even if $\tau$ is not improved thanks to a factor of 100 lower temperatures afforded by adiabatic nuclear demagnetization methods. 

In our previous work, we have developed a setup allowing for stable levitation and motion control of superconducting objects at single kelvin temperatures \cite{Arrayas_Trueba_Uriarte_Zmeev_2021,Arrayas_Bettsworth_Haley_Schanen_Trueba_Uriarte_Zavjalov_Zmeev_2023}. Going a step further, we have constructed a similar setup compatible with advanced wet dilution refrigerators \cite{bradley1994simple, cousins1999advanced}.  In this configuration, we can levitate macroscopic objects at temperatures below 3\,mK in a highly isolated environment, having orders of magnitude better vibrational noise compared to dry, pulsed tube based cooling systems \cite{schmoranzer2019cryogenic}. Our refrigerators are designed for adiabatic nuclear demagnetisation experiments and as such require ultimate vibration isolation. Performing direct measurements of the vibration spectra of the refrigerator can prove challenging. However, vibrations are usually the dominating source of heat leak to the refrigerant as they induce eddy currents due to motion in a magnetic field. The measured heat leak to the refrigerant, therefore, serves as a direct integral characteristic of the effectiveness of the vibration isolation.  As a figure of merit, the lowest lattice temperature of copper cooled in a similar wet refrigerator was recorded at 1.5\,$\mu$K. Such refrigerators have demonstrated week-long hold time below 20\,$\mu$K, with the residual heat leak comparable to an estimated heating rate due to interactions with cosmic rays and ambient radiation~\cite{Pickett2000,autti2024quest}. For comparison, record temperatures of advanced nuclear demagnetisation cryostats based on dry dilution refrigerators remain at $90~\mu$K, with hold times of $\sim10$\,h below 100\,$\mu$K. The vibrations-induced heat leak was found to dominate~\cite{Yan2021}. The heat leak due to the background radiation and cosmic rays can be, in principle, further diminished by careful choice of construction materials of the refrigerator and by moving the experiment underground.


\section{Results}

\begin{figure*}[t]
\includegraphics[width=\textwidth]{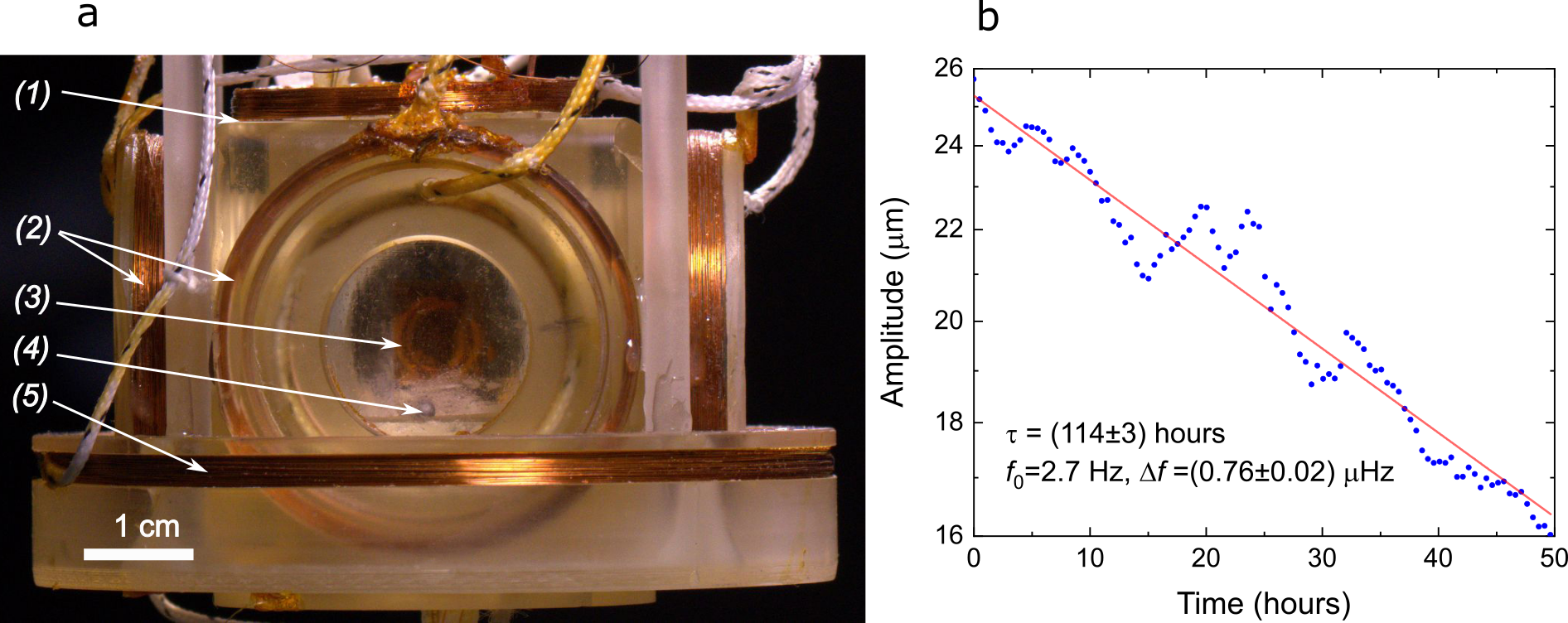}
\caption{(a) Dilution refrigerator compatible setup for  levitation of superconducting objects. It allows precise position manipulation and detection in the amplitude range from $\mu$m up to cm. The 2\,mm lead-plated ball (4) is levitated in the magnetic field created by two coplanar and coaxial superconducting NbTi coils in a copper matrix (5). The ball's position can be manipulated in the horizontal plane using two pairs of lateral gradient coils made from the same material (2). Nominally, the ball is levitated in the plane defined by the axes of these coils. The position of the levitating ball is determined by measuring the radio-frequency signal dependent on the position of the ball between the transmitter coil  (1) and a pair of mutually orthogonal receiver coils (3). The ball is inside a closed cell that can be evacuated via a capillary. See~\cite{Suppl} for more details. (b) Decay of free lateral oscillations of the levitator in vacuum at the base temperature of the dilution refrigerator (5\,mK). The resonant frequency $f_0$ was 2.7\,Hz and the ring-down time was $(114\pm3)$ hours corresponding to the resonant width of $\Delta f=(0.76\pm0.02)\,\mu$Hz.}
\label{fig:setupanddecay}
\end{figure*}

The levitation system follows the design proposed in \cite{Arrayas_Trueba_Uriarte_Zmeev_2021}, where a superconducting object is magnetically levitated by two concentric superconducting coils energised with opposing currents, creating a levitation potential which is almost flat in the horizontal plane. For better accessibility of theoretical description and fabrication process optimization, we have decided for the levitated object to be a sphere. We chose type I superconductor (lead) to avoid problems related to flux line pinning/depinning within type II superconductors and the resulting non-reproducibility of its motion. The flux line motion within an oscillating type II superconductor increases dissipation as well~\cite{schilling2007design}. The system is shown in Fig.~\ref{fig:setupanddecay}a. The gravito-magnetic levitation potential can be adjusted by varying the currents in the levitation coils. Levitated object can also be maneuvered in the horizontal plane by two pairs of lateral coils as the application of a lateral field gradient shifts the position of the equilibrium. This enables large scale motion with high speeds and amplitudes larger than the object's size, following linear or circular trajectory. This motion is required for superfluid dynamics studies, results of which we will publish elsewhere. At the same time, using the lateral gradient coils, we can impose a shift of the equilibrium position in the horizontal plane in order to tune the natural frequency of small-amplitude free oscillations.

Special care must be taken when constructing the detection system for the position resolution of the levitator with sufficient sensitivity in the whole displacement range from $\mu$m to cm. A SQUID-based detection system, such as realised in~\cite{fuchs2024measuring,Hofer_Gross_Higgins_Huebl_Kieler_Kleiner_Koelle_Schmidt_Slater_Trupke_Uhl_Weimann_Wieczorek_Aspelmeyer_2023}, allows for resolution of small-amplitude motion, however it is not appropriate for registering large-displacement motion controlled by relatively strong magnetic field pulses required in our system. Therefore, we chose an optomechanical detection scheme based on measuring the radio-frequency transmission between two coils in the space containing  the diamagnetic ball. This scheme offers a suitable compromise between precision and dynamic range of position measurements. 

We start the experiment by ramping the current in the two levitation coils to the pre-calculated levels and persist the currents in the coils using heat-activated persistent switches. We then press the ball against the vertical wall of the cell by ramping current in the lateral coils, creating a horizontal field gradient. This allows us to damp out any residual oscillations and to bring the ball to a quiescent state. We then slowly move the ball away from the wall with a quasi-uniform speed of $\sim10\,\mu$m\,s$^{-1}$ to a desired position~\cite{Zmeev_2014}. The position of the ball in the horizontal plane defines its eigen frequency. We choose the resonant frequency to be away from the pre-measured residual vibration maxima of the dilution refrigerator. We then drive a small current at the eigen frequency of the ball for a few periods to set the ball in motion in the desired direction. We then persist the gradient coils using the same type of persistent switch. We measure the position of the ball as a function of time using the signal received from the LC-circuit containing one of the receiver coils and calibrated against the position of the same ball observed in an optical cryostat at 1.5\,K. We can also calibrate the position by recording the signal against the position while slowly changing the field gradient produced by the lateral coils. The position can be calculated by finding the equilibrium position of the ball based on its measured mass and radius for the given currents in the control coils. In optical experiments, we found a very good match between the observed and calculated positions. We have also performed 3D FEM simulations of the detection system~\cite{Suppl}.


The usual frequencies of measured free oscillations were in the single Hz range for the horizontal mode and about 10\,Hz for the vertical mode. The evolution of the detection signal amplitude was then measured in five-minute-long blocks every hour for multiple days. Each data block was processed using Fourier spectral analysis to obtain the amplitude of the studied mode. In all presented data we have observed a clean, well-separated peak of the induced mode. 


Fig.~\ref{fig:setupanddecay}b shows free oscillations of the levitated superconducting ball in cryogenic vacuum at 5\,mK, revealing a ring-down time longer than 400\,ks (114 hours). This result represents extremely low losses in the system and, to our knowledge, the record value for coherent oscillations of a superconductor in a magnetic field. 

\begin{figure*}[t]
\includegraphics[width=0.8\textwidth]{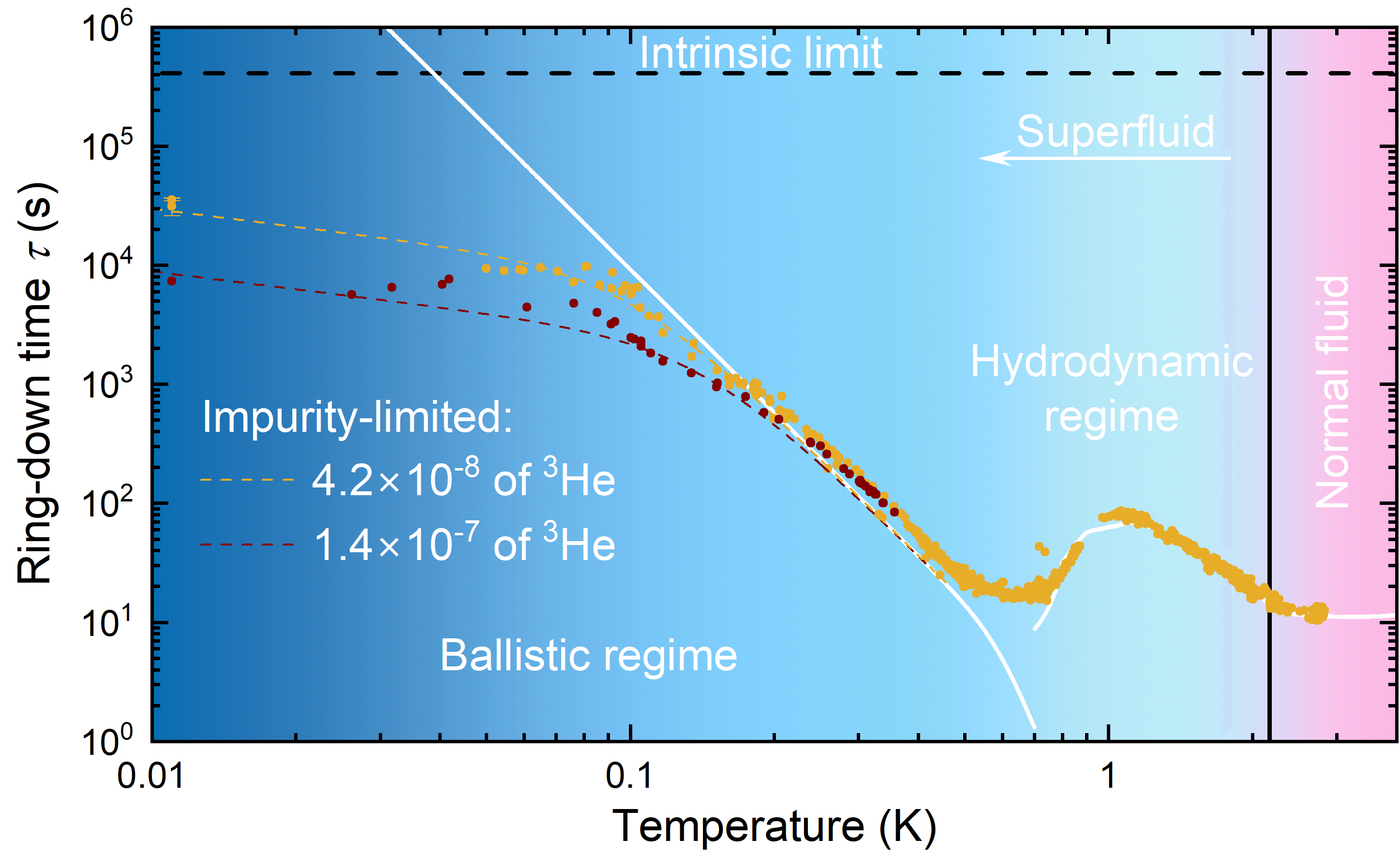}
\caption{Dissipation of the oscillator in liquid helium-4. The ring-down time is determined by the properties of the surrounding fluid. The predicted behaviour based on the mass and the radius of the ball is shown by the white line. As the temperature decreases, the dissipation is first governed by viscous drag due to the normal component~\cite{donnelly1998observed}; then ballistic excitations (mainly phonons)~\cite{Jager_Schuderer_Schoepe_1995a, Jager_Schuderer_Schoepe_1995b}; and, finally, the ballistic drag of $^3$He impurities in the superfluid (coloured dashed lines)~\cite{Suppl}. The intrinsic vacuum limit (dashed black line) would correspond to a damping due to $\sim10^{-9}$ of $^3$He at 40\,mK. The presence of $^3$He impurities at the level of $10^{-8}$ has not been detected before with any other type of oscillator. }
\label{fig:helium_decay}
\end{figure*}

We have repeated this experiment in the cell filled with liquid helium and we can observe the expected behaviour of the decay time constant of the oscillations of the ball over many orders of magnitude as the temperature is varied~\cite{Andronikashvili_1946,Jager_Schuderer_Schoepe_1995a,Jager_Schuderer_Schoepe_1995b}, as shown in Fig.~\ref{fig:helium_decay}. Remarkably, we can see the dissipation due to $\sim3\times10^{-8}$ of $^3$He impurities~\cite{Suppl}. We have used nominally isotopically pure $^4$He for our experiment~\cite{hendry1987continuous}, however the fill lines and the gas-handling system have been used for $^3$He experiments previously and the $^4$He must have become contaminated by a trace amount of $^3$He. We have later added a measured amount of $10^{-7}$ of $^3$He impurities, and the ring-down time due to drag by impurities has decreased according with the expectation.  No other existing mechanical oscillator would be sensitive for a small change in the dissipation rate due to drag from this small amount of impurities.

In linear regime, the drag force exerted on the ball is 
\begin{equation}
    F_\mathrm{D}=\frac{2M}{\tau}V, 
\end{equation}
 where $V$ is the velocity of the ball and $M=6.33$\,mg is its mass. In the current setup, we can resolve $\sim1\,\mu m$ displacements of the ball, corresponding to a r.m.s. velocity $V\sim10\,\mu m\,s^{-1}$ at our typical frequencies of 3\,Hz. For $\tau=4\times10^{4}$\,s, measured in the superfluid, the drag force exerted by the $^3$He impurities is $F_\mathrm{D}\sim3\times10^{-15}$N=3\,fN. The intrinsic limit for $\tau=4.1\times10^5$\,s would correspond to a fraction of a femtonewton exerted on a milligram-scale mass. This is on par with the force sensitivity demonstrated recently for a silicon nanoparticle of attogram mass~\cite{kostarev2025attonewton} and a sub-milligram levitating mass~\cite{fuchs2024measuring}. 
 
 We note that our position detection system is not designed for sub-micrometer measurements. A position detection system based on a SQUID fluxmeter would offer a 5-6 orders of magnitude improvement on the  position resolution~\cite{fuchs2024measuring, Hofer_Gross_Higgins_Huebl_Kieler_Kleiner_Koelle_Schmidt_Slater_Trupke_Uhl_Weimann_Wieczorek_Aspelmeyer_2023, pratt2011interplay}, with the force sensitivity increasing by the same amount and potentially offering zeptonewton-scale measurements with a milligram mass.


\section{Discussion}

In summary, implementing a levitator made from type I superconductor, removing the influence of residual gas, and using an advanced dilution refrigerator with vibration isolation has allowed us to achieve extremely low dissipation levels in a levitating superconducting oscillator, equivalent to $(0.76\pm0.02)\,\mu$Hz resonant linewidth.

The level of vacuum dissipation rate in the current setup should allow our probe to be sensitive to $<10^{-9}$ level of $^3$He impurities in superfluid $^4$He. A similar but lighter probe coupled with a SQUID-based fluxmeter detection system would be able to be sensitive to single vortices in the superfluid and would enable detailed studies of the Kelvin wave cascade on a single vortex~\cite{eltsov2020amplitude}. In $^3$He-B, a similar levitator would be an extremely sensitive probe of quasiparticle excitations as used in bolometers for dark matter search~\cite{quest2024quest}: resonators currently employed have coherence times of only $\sim10^2$\,s, significantly limiting the sensitivity. The built-in ability to detune the oscillator from the vibrational noise bands of the refrigerator is also a significant advantage.




Even though an order of magnitude lower dissipation rates have been achieved with dielectric and diamagnetic particles~\cite{leng2021mechanical, dania2024ultrahigh} at elevated temperatures, our approach offers a straightforward pathway to lowering the $T/\tau$ parameter by more than two orders of magnitude further in nuclear demagnetisation refrigerators. The levitator is highly tunable and is designed for microkelvin experiments in a highly vibration-isolated environment.


Removing further potential dissipation factors, such as normal metal in the cladding of the control coils, would allow us to  increase the coherence time further. Increasing the precision of the position sensing by implementing a SQUID-based fluxmeter, will allow us to manipulate the levitator’s motion on the pico- and femtometer-scale and to implement feedback cooling of the chosen mode of oscillations, potentially resulting in cooling to the quantum mechanical ground state. Decreasing the bath temperature further, down to below 100\,$\mu$K by using readily available adiabatic nuclear demagnetisation, will enable us to bring the $T/\tau$ parameter to below 10$^{-10}$\,K\,s$^{-1}$. At this level, quantum measurements with the levitator might become possible, opening access to experiments probing the interface between classical and quantum mechanics, such as quantumness of gravity.

\section{Acknowledgements}

This work has been funded by UKRI (EP/X004597/1), GA\v CR (24-12601O), Spanish Ministry of Science PID2022-139524NB-I00, and URJC Programa Propio 2025/00014/043. We acknowledge expert technical support of  Mechanical workshop machinists at Lancaster Physics R.~Grimshaw and J.~Meadowcroft.

\bibliography{articlebib}
\clearpage

\section*{Supplementary Information}
\renewcommand{\thesection}{\Alph{section}}
\setcounter{section}{0}
\section*{A. Levitation setup and helium cell}

We have performed experiments levitating a 2\,mm diameter lead-covered Delrin ball measuring its horizontal modes at Hz frequencies and its vertical mode at about 10 Hz. 

\begin{figure}[h]
\includegraphics[width=0.5\textwidth]{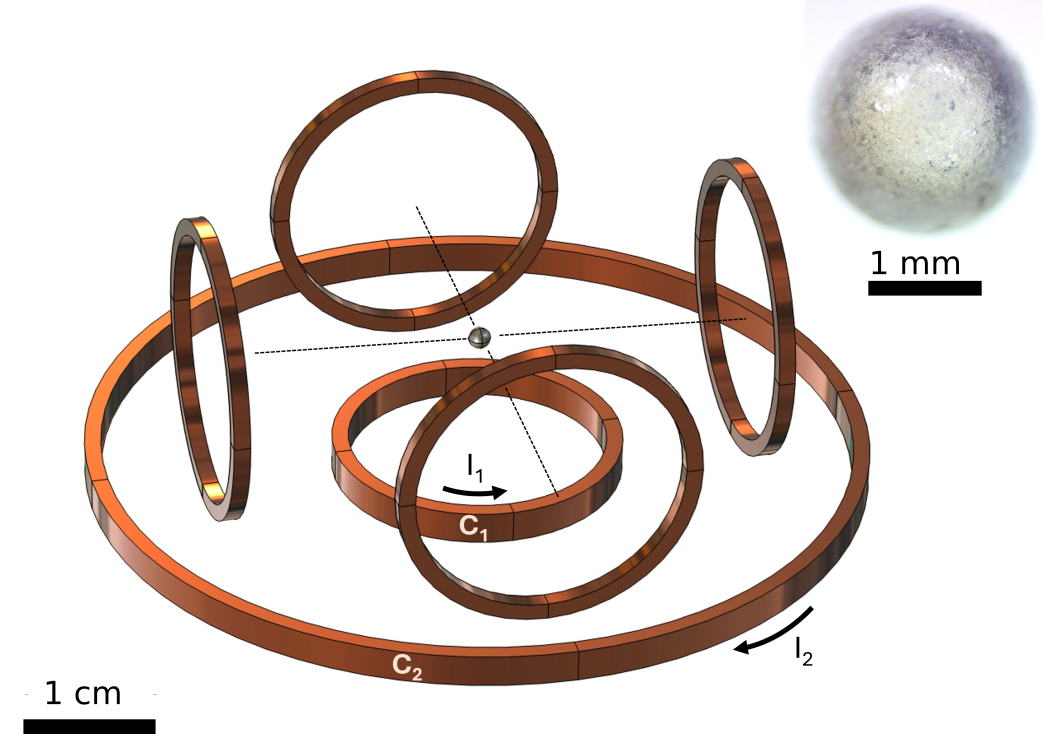}
\caption{Schematic of the levitation control system~\cite{Arrayas_Trueba_Uriarte_Zmeev_2021}. It contains two levitation coils (C1 and C2), carrying opposing and equal currents $I_1$ and $I_2$. There are two pairs of lateral displacement control coils. The levitation potential is designed to be almost flat when the ball is levitated in the plane of the axes of the lateral coils. All coils have heat-activated persistent switches. The inset shows an optical microscopy picture of the fabricated lead shell of 2\,mm diameter. Interior of the shell is comprised of silver-plated Delrin ball~\cite{Arrayas_Bettsworth_Haley_Schanen_Trueba_Uriarte_Zavjalov_Zmeev_2023}.}
\label{fig:sphere}
\end{figure}

One of the goals of our experiment is to move the object at high velocities within quantum fluids, requiring fast accelerations and decelerations~\cite{zmeev2014method, zmeev2015dissipation, Bradley_Fisher_Guenault_Haley_Lawson_Pickett_Schanen_Skyba_Tsepelin_Zmeev_2016,Autti_Haley_Jennings_Pickett_Poole_Schanen_Soldatov_Tsepelin_Vonka_Zavjalov_Zmeev_2023, autti2020fundamental}. For this reason, instead of a solid lead ball, we used a much lighter superconducting shell manufactured by electroplating lead~\cite{Abdel-Aziz_El-Zomrawy_El-Sabbah_Ghayad_2022} onto silver-coated \cite{Arrayas_Bettsworth_Haley_Schanen_Trueba_Uriarte_Zavjalov_Zmeev_2023, Tollens_1882} Delrin balls, see Fig.~\ref{fig:sphere}. Lead was chosen for its relatively high critical field and temperature, which is vital whenever position manipulation pulses are applied. The typical thickness of the resulting lead shell has been estimated at 15\,$\mu$m which is well above the London penetration depth \cite{Lock_Bragg_1951, Greytak_Wernick_1964, Gasparovic_McLean_1970, Egloff_Raychaudhuri_Rinderer_1983}, with the surface roughness of the order of $\mu$m.  

The crucial part of the experiment is mitigation of all sources of parasitic dissipation present in the system. For this reason, the bulk of the setup was fabricated from Araldite, having good rigidity and low levels of radioactive background, which is known to be a possible limiting factor for ultra-low temperature experiments \cite{autti2024quest}. We have chosen a threaded modular assembly of the system's body and coil formers reducing any misalignment and relative displacement upon cooldown. All levitating coils and horizontal position control coils are wound from Cu-matrix multi-filament NbTi superconducting wires connected in parallel with custom-made superconducting persistent switches allowing isolation from any power source electrical noise. Once the levitator is set in motion using a small periodic pulse of control current, all control coils are persisted. We did not find any change in the trapped current after a few months of experiments. Indeed, current stability better than $10^{-11}$/h was demonstrated in similar systems~\cite{van1999ultrastable}. This result ensures stability of the levitation potential. We have made extra steps to exclude any large quantities of normal metal in the vicinity of the oscillator to avoid eddy current losses. Direct optimization of the setup lies within the rewinding of the coils with NbTi wire free from the normal-metal matrix. In addition, we have installed partial Nb shielding  closely around the levitating setup wherever possible to exclude spontaneous trapping of Earth's magnetic field by surface defects at the transition temperature of the levitator. Since the shielding was retro-fitted, it was impossible to fully enclose the experiment with a superconducting shield. Finally, the experiment was performed in a shielded room and in custom-made wet dilution refrigerators with advanced vibration isolation~\cite{bradley1994simple, cousins1999advanced}. 

\begin{figure}[t]
\includegraphics[width=0.4\textwidth]{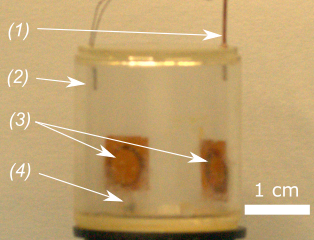}
\caption{Experimental cell machined from Araldyte and glued together with Stycast 1266 can be evacuated and filled via the copper capillary (1). It contains a quartz tuning fork viscometer (2)~\cite{blaauwgeers2007quartz} and  the lead-coated ball (4). The cell was inserted in the center of the levitation  setup.  Bare NbTi coils (3), being part of LC detection circuits, are rigidly glued to the cell walls.}
\label{fig:cellphoto}
\end{figure}

The experimental volume, i.e. a fully separated cylindrical Araldite cell, containing the  lead-coated ball shown in Fig.~\ref{fig:cellphoto}, was thoroughly evacuated using a diffusion oil pump for multiple days, prior to cooldown, in order to minimise the amount of any gaseous impurities. The whole system was then cooled to base temperature of the dilution refrigerator, about 5\,mK, and left to thermally equalize for a week. The temperature of the refrigerator continued to decrease during this time due to decreasing parasitic heat leak from relaxing two-level systems within the plastic parts~\cite{pobell2007matter}. The superconducting ball was then levitated to 7\,mm above the cell floor by ramping up opposing DC currents ($\approx$ 4\,A each) in the levitation coils using a computer-controlled multichannel current supply. The equilibrium position of the ball in the 3D space was set via fine-tuning the levitation currents and the horizontal control coils currents, supplied by Kepco BOP-20M power supplies controlled by a DAQ system, to choose natural frequencies of vertical/horizontal oscillations. This allowed us to avoid any natural frequencies of vibrational noise in the measured spectrum of our dilution refrigerator. 

\section*{B. Position detection system and its calibration}

In this section we present the principles behind our detection system based on the perturbation of a magnetic field by a diamagnetic sphere and its practical realization. In a conceptually simpler case, the system consists of two coaxial coils facing each other, where one coil is driven by a time-dependent current while the second coil acts as a receiver. The presence of the diamagnetic sphere modifies the magnetic flux penetrating the receiver coil due to flux exclusion, leading to a measurable variation in the induced signal. This variation depends on the position of the sphere and can therefore be used as a position-sensing mechanism.
	
	 The model is formulated within the quasi-magnetostatic (QMS) regime and implemented using the finite element method (FEM) under axial symmetry. Particular attention is given to the treatment of the superconducting coils, the diamagnetic response of the sphere, and the inclusion of displacement current effects within the adopted approximation.
	

	\subsection*{Problem Description and Mathematical Model}

	The detection system comprises two circular coils: a primary transmitting coil and a secondary receiving coil. These may vary in radii and number of turns, and need to be positioned far enough from the trajectory of the sphere moving in a fluid as not to perturb the flow field. 
	
	Our superconducting type I sphere may be regarded (at least considering its magnetic field distortion under perfect conditions of the applied field) as a perfectly diamagnetic spherical object of radius $r$ which is positioned along the space delimited by the coils and is allowed to move through it. Due to its ideal diamagnetic behavior, the sphere expels magnetic flux from its interior, thereby perturbing the magnetic field distribution generated by the transmitting coil.
	
	The idea of the detection system consists in applying a time-dependent current in the transmitting coil:
	\begin{equation}
		I(t) = I_0 \cos(\omega t),
	\end{equation}
	so the time-varying magnetic field generated by the transmitting coil induces a voltage in the receiving coil through magnetic coupling. In the absence of the sphere, the induced voltage is determined by the mutual inductance between the coils. However, the presence of the diamagnetic sphere modifies the magnetic flux linking the receiving coil, thereby altering the effective inductive coupling which can be measured to extract the position of the sphere. Note that this principle does not require for both coils to be coaxial.

	For simplification reasons, let us assume both detection coils (transmitter and receiver) are placed one in front of the other (coaxial), sharing their axes and being separated by a distance $d_z$. We also assume that the axis of both coils corresponds to $z$ direction. Let $\Phi(t)$ denote the magnetic flux through the receiving coil. The effective inductance can be defined as
	\begin{equation}
		L_{\text{eff}}(t) = \frac{\Phi(t)}{I(t)}.
	\end{equation}
	Due to the flux expulsion imposed by the superconducting sphere, the magnetic flux through the receiving coil decreases as the sphere approaches it. Consequently, the effective inductance becomes a function of the sphere position $z$ along the axis
	\begin{equation}
		L_{\text{eff}} = L_{\text{eff}}(z),
	\end{equation}
	with
	\begin{equation}
		\frac{dL_{\text{eff}}}{dz} < 0.
	\end{equation}
	  In practical implementations, the receiving coil is typically part of an $LC$ resonant circuit (where $C$ represents the total capacitance of the coil circuit, including the parasitic capacitance). The resonance frequency is given by
	\begin{equation}
		f = \frac{1}{2\pi \sqrt{L_{\text{eff}}(z) C}}.
	\end{equation}
	Since $L_{\text{eff}}$ depends on the sphere position, the resonance frequency becomes position-dependent
	\begin{equation}
		f = f(z).
	\end{equation}
	Hence, a displacement of the sphere therefore results in a measurable shift in the resonance frequency. Also, the voltage induced in the receiving coil is governed by Faraday's law
	\begin{equation}
		V_{\text{ind}}(t) = -\frac{d\Phi(z)}{dt} = \int_{\Sigma}^{} B_z\cdot dS,
	\end{equation}
	where $\Sigma$ represents the surface of the coil. Using the definition of inductance, this can be expressed as
	\begin{equation}
		V_{\text{ind}}(t) = -\frac{d}{dt}\left( L_{\text{eff}}(z) I(t) \right).
	\end{equation}
	Assuming that the motion of the sphere is slow compared to the excitation frequency (quasi-static position), $L_{\text{eff}}(z)$ can be treated as constant over one oscillation period, yielding
	\begin{equation}
		V_{\text{ind}}(t) \approx -L_{\text{eff}}(z) \frac{dI(t)}{dt}.
	\end{equation}
	Which is also a measurable quantity to detect the position of the sphere.
	
	\subsection*{Mathematical Model and FEM Formulation}
	
	The system is modeled by solving the Ampère equation
	\begin{equation}
		\nabla \times \mathbf{H} = \mathbf{J},
	\end{equation}
	where $\mathbf{H}$ is the magnetic field, $\mathbf{J}$ is the imposed current density in the coils. The magnetic vector potential $\mathbf{A}$ in the Coulomb gauge is introduced such that
	\begin{equation}
		\mathbf{B} = \nabla \times \mathbf{A}.
	\end{equation}
	Assuming linear material properties, the constitutive relations are given by
	\begin{equation}
		\mathbf{B} = \mu \mathbf{H}.
	\end{equation}
	The problem is formulated in terms of the magnetic vector potential $\mathbf{A}$ and solved over the entire computational domain. The coils are modeled as ideal superconductors, and therefore the current density is imposed directly as
	\begin{equation}
		\mathbf{J} = \frac{N \textbf{I}}{S_{\text{cond}}},
	\end{equation}
	where $N$ is the number of turns, $I$ is the applied current, and $S_{\text{cond}}$ is the total cross-sectional area of the conductor. On the surface of the diamagnetic sphere, a magnetic flux exclusion condition is imposed
	\begin{equation}
		\mathbf{n} \times \mathbf{A} = 0,
	\end{equation}
	which enforces  vanishing of the tangential component of the magnetic vector potential and ensures magnetic field expulsion from the sphere. Zero initial conditions are imposed throughout the domain, such that both the magnetic vector potential and the electric scalar potential vanish at the initial time
	\begin{equation}
		\mathbf{A}(t=0) = 0.
	\end{equation}
	The problem is solved numerically using FEM. Quadratic shape functions are employed for spatial discretization, and time integration is performed using an implicit Backward Euler scheme. The magnetic field is computed from the curl of the magnetic vector potential. The magnetic flux through the receiver coil is obtained by integrating the magnetic field over the middle cross-sectional area of the coil and multiplying by the number of turns.

	\subsection*{Simulation parameters and tuning}
	
	The specific design tested consists of two coaxial, horizontally aligned facing coils and a diamagnetic sphere positioned between them. The sphere has a radius of $r=1$\,mm, while the separation between the coils is $d_z=2.27$\,cm.
	
	The receiver coil is composed of $N_r = 60$ turns and has a radius of $r_r = 3$ mm, while the transmitter coil consists of $N_t = 100$ turns with a radius of $r_t = 12.5$ mm. The total conductor cross-sectional area is $0.012$\,cm$^2$ for the receiver and $0.022$\,cm$^2$ for the transmitter, with an individual conductor cross-section of $7 \times 10^{-9}$\,m$^2$. The choice of parameters for the coils is not arbitrary. Both coils (transmission and detection) must be fitted into the cavity located within the lateral control coils, so the maximum size of the former is determined by the size of the latter. 
	
	The surrounding medium corresponds to liquid helium, characterized by a relative permittivity $\varepsilon_r = 1.048$, relative permeability $\mu_r = 1$, and electrical conductivity $\sigma = 10^{-13}$\,S/m.
	
	The total capacitance coupled to the receptor is estimated from the experimentally observed resonance frequency and inductance. Assuming a resonance frequency of approximately $f_0 \approx 1.6$ MHz and an inductance of $L \approx 21\,\mu$H, the capacitance is obtained from the standard LC resonance relation
	\begin{equation}
		f_0 = \frac{1}{2\pi \sqrt{LC}},
	\end{equation}
	which yields an approximate capacitance of $C \approx 470$\,pF.
	
	Two independent simulation approaches are carried out. In the first approach, the receiver coil is excited with a sinusoidal current, and a time-domain simulation of duration $1\,\mu$s is performed for each sphere position. The magnetic flux is computed, and the effective inductance is obtained as the ratio between flux and current
	\begin{equation}
		L_{\text{eff}}(t) = \frac{\Phi(t)}{I(t)}.
	\end{equation}
	Using the previously estimated capacitance, the corresponding resonance frequency is then recalculated for different positions of the sphere, starting at 1.9\,cm to 0.2\,cm away from the receiver, measured with respect to the middle cross section of the receiver coil, allowing the frequency shift to be determined as a function of position.
	
	In the second approach, the transmitter coil is excited with a sinusoidal current. This current is extracted from the experiments by measuring the induced voltage in the receiver in the absence of the sphere, for a given voltage applied in the transmitter. The current takes the form
	\begin{equation}
		I(t) = 0.035 \sin(2\pi \cdot 1.6 \times 10^6\ t).
	\end{equation}
	For each sphere position, a time-domain simulation of $1\,\mu$s is performed, and the induced voltage in the receiver coil is computed from the time variation of the magnetic flux.
	
	All simulations are carried out with a maximum time step of $\Delta t_{\max} = 10^{-8}$ s using an implicit Backward Euler scheme.
	
	The experimental results were obtained by placing the sphere into different positions along the detection pair of coils axis. To measure the position, optical measurement techniques were employed. The transmitter is supplied with a generator and the  induced signal is picked up and amplified using a lock-in amplifier. The resonance frequency was first obtained performing a frequency sweep analysis  and then the lock-in is tuned in that frequency to measure the induced voltage in the receiver.
	
\begin{figure}[t!]
		\centering
		\begin{subfigure}[b]{0.48\textwidth}
			\centering
			\includegraphics[width=\textwidth]{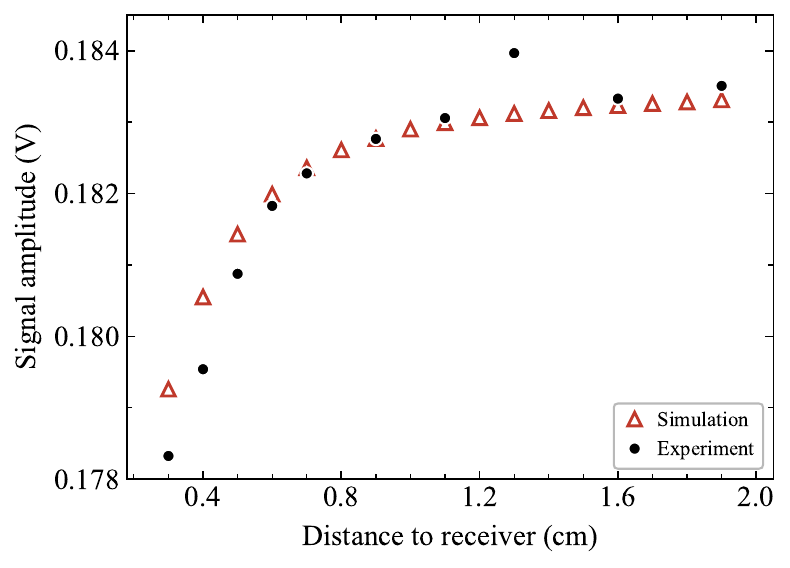}
			\caption{}
			\label{fig:amplitude}
		\end{subfigure}
		\hfill
		\begin{subfigure}[b]{0.48\textwidth}
			\centering
			\includegraphics[width=\textwidth]{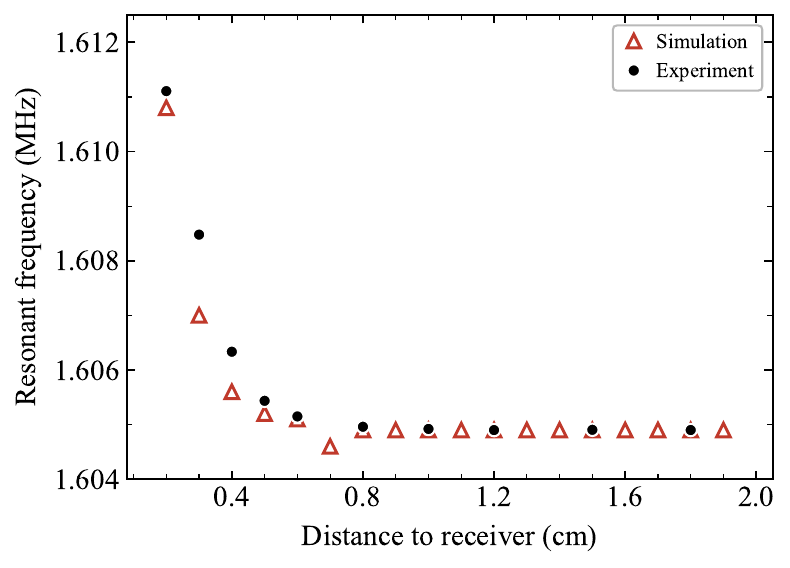}
			\caption{}
			\label{fig:frequency}
		\end{subfigure}
		\caption{Comparison of experimental and simulation results as a function of distance to the receiver showing a very good agreement. (a)~Signal amplitude. (b)~Resonant frequency.}
		\label{fig:comparison}
	\end{figure}

	Finally, the numerical results obtained from both simulations were compared with experimental measurements performed under equivalent conditions. The results are shown in Fig.~\ref{fig:comparison}.

	The presented coaxial detection system has a substantial problem that can be noticed in both graphs of Fig.~\ref{fig:comparison}. Although it is possible to see the change in the position of the sphere in both graphs, as it moves away from the detection coil, the change in the signal becomes smaller, with the variation being minimal once the sphere has moved beyond the halfway point between the two coils on the same axis. The reason for this is clear: as the sphere moves away from the receiver, the magnetic flux deflected by the sphere away from the receiver is reduced, and therefore the change in the voltage and inductance of the receiving coil significantly decreases. 
    
	\begin{figure}[t]
		\centering
		\includegraphics[width=0.63\textwidth]{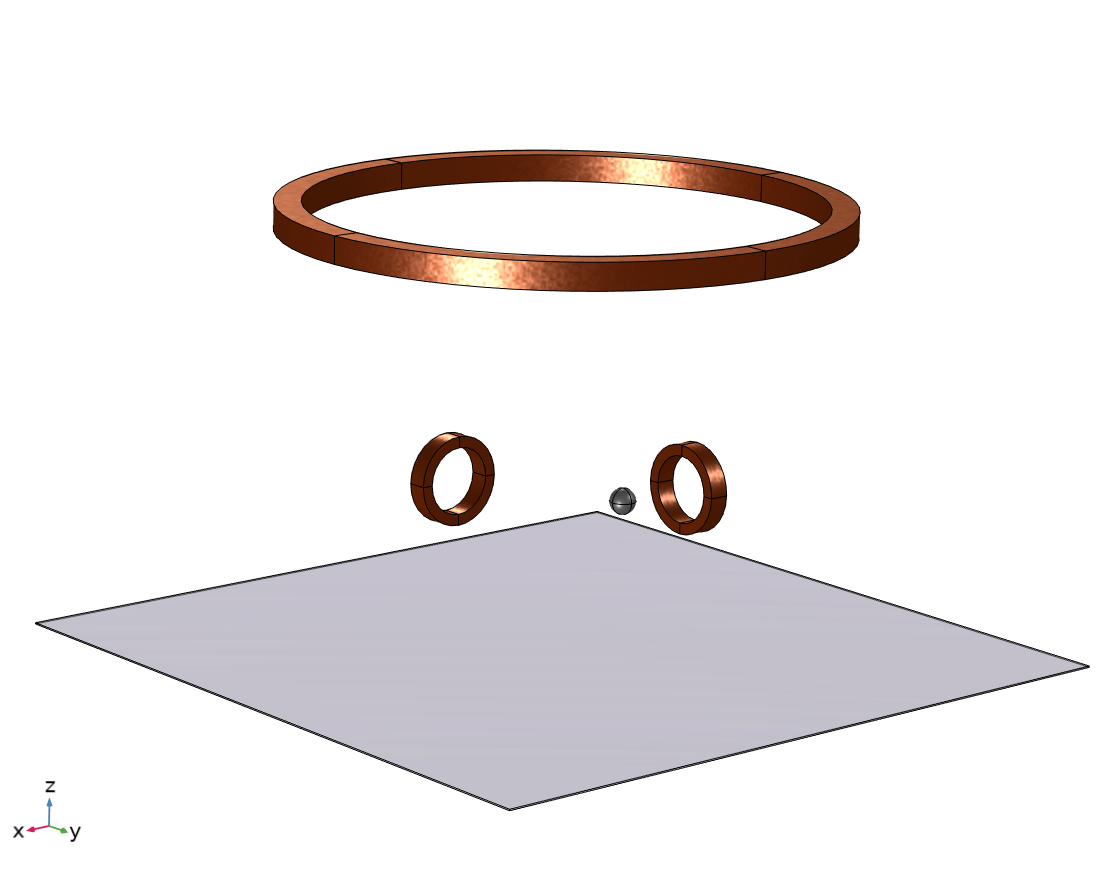}
		\caption{Detection system sketch. The transmission coil is located outside the cell at a vertical distance of 2.2\,cm above the levitation plane. The two additional coils located below represent the receiver coils, with a 6\,mm radius. Their axes are placed in the levitation plane. Note that the simulations described in the text have been carried out with only one receiver coil present.}
		\label{fig:det_sketch}
	\end{figure}

	To overcome these issues, we have modified the setup by including two orthogonal receiver coils (along each horizontal direction placed inside two of the control lateral coils) and by implementing a single transmission coil, located on the top of the cell. A sketch of such setup is shown in Fig.~\ref{fig:det_sketch}. This design offers numerous advantages over the previous one. First, the levitation cell configuration has plenty of space on the upper surface, making it possible to position a large transmission coil. Secondly, the top placement makes it possible to scale the model without increasing the distance between the transmission and receiver coils (the distance from the transmission coil to the levitation plane, where the detection coils axes cross can be reduced in our geometry to 2.2\,cm). Finally, it is clear that using a single transmission coil fed by a sum of currents at different frequencies reduces the complexity of the system and allows for the introduction of additional detection coils without increasing the number of transmission coils. 
			
\begin{figure}[th!]
		\centering
		\includegraphics[width=0.95\textwidth]{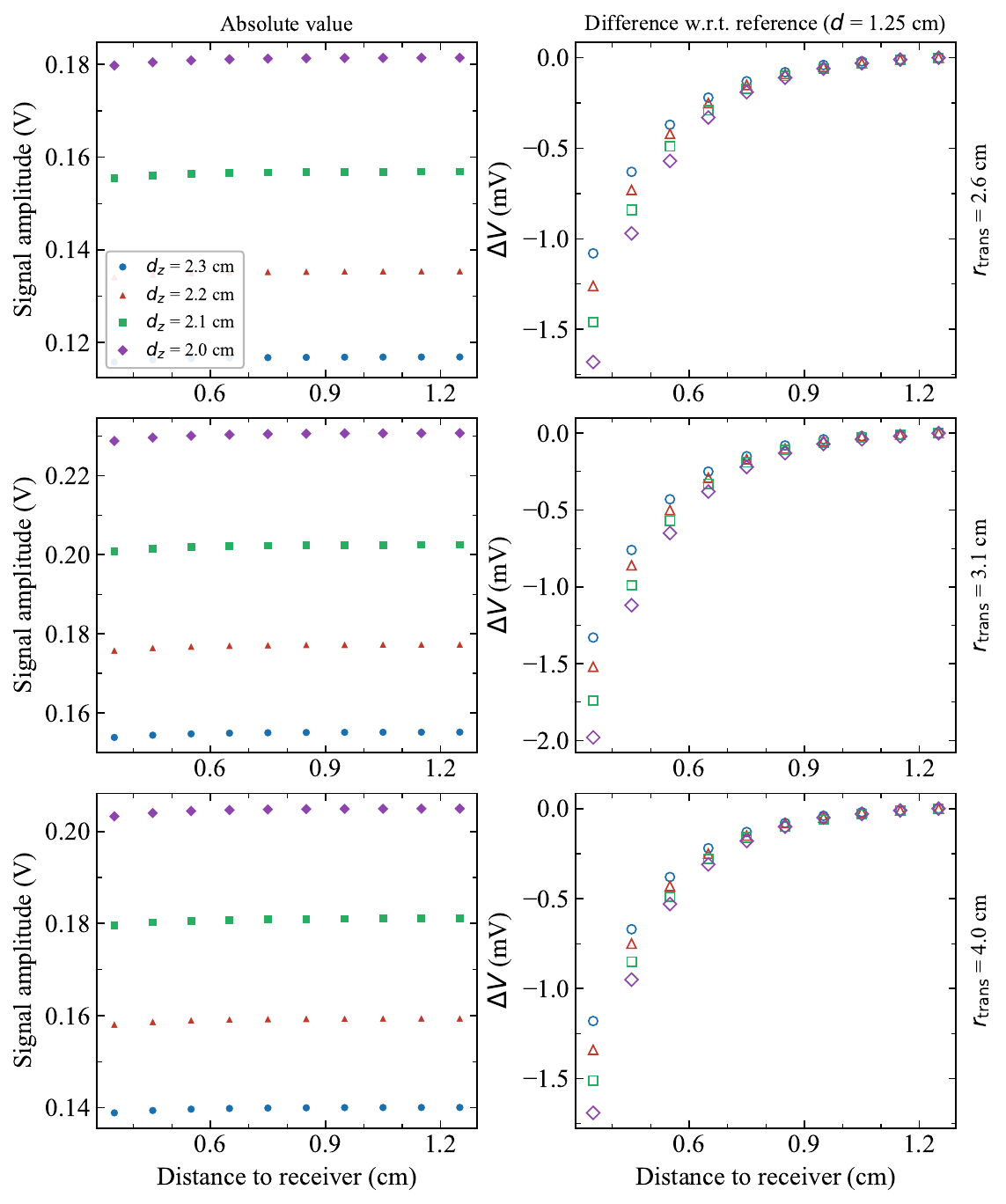}
		\caption{Signal amplitude as a function of distance to the receiver for three values of the transmitting coil radius ($r_{\mathrm{trans}} = 2.6,\,3.1,\,4.0\,\mbox{cm}$) and four vertical separation distances ($d_z = 2.0,\,2.1,\,2.2,\,2.3\,\mbox{cm}$) measured from the transmitting coil plane to the levitation plane. Left column: absolute signal amplitude, where the dominant variation is due to the vertical separation. Right column: amplitude difference $\Delta V$ with respect to the reference value at  distance to the receiver coil $d = 1.25\,\mbox{cm}$. Each row shares independent vertical scales to enhance the visibility of the trends within each panel.}
		\label{fig:det_3D}
\end{figure}

	This new design allows us to modify the parameters of the transmission coil, since we are no longer restricted by the size of the lateral control coils. To achieve the highest detection signal, we have performed similar simulations as the ones presented in the previous subsection. In this case, we could not use axisymmetric models. Instead, we have implemented fully 3D FEM simulations of MQS equations. The receiver coils still remain with 6\,mm radii and 60 turns each. However, we have performed simulations varying the radius of the transmission coil from $r_{\mathrm{trans}} = 2.6 \, \mbox{cm}$ to $r_{\mathrm{trans}} = 4\,\mbox{cm}$. The vertical separation has been also varied from $d_{\mathrm{z}} = 2.0\, \mbox{cm}$ to $d_{\mathrm{z}} = 2.3\, \mbox{cm}$. This vertical distance is measured as the distance from the transmitter coil to the levitation plane. The simulations have been performed placing the sphere in different positions along the axis of one of the receiver coils, and measuring the induced voltage. These results are represented in Fig.~\ref{fig:det_3D}.

    \clearpage




\section*{C. Linear damping of a sphere oscillating in superfluid $^4$He}

We have measured the dissipation of the sphere immersed in liquid helium in the linear (laminar) regime of small-amplitude oscillations in a wide range of temperatures. We will publish the results of the measurements in the turbulent regime elsewhere.

In the hydrodynamic regime, the decay time of oscillations of a ball of mass $M=6.33$\,mg and radius $r=1.00$\,mm, oscillating in liquid helium is given by Stokes' formula

\begin{equation}
    \tau_{\mathrm{hydr}} = \frac{M} {3\pi \eta_n r},
    \label{eq:Stokes}
\end{equation}
where $\eta_n$ is the viscosity of the normal component~\cite{donnelly1998observed}.
When calculating the radius, we account for the estimated thermal contraction of the ball at a level $1.5\%$, typical of polymers~\cite{pobell2007matter}.
For our relatively low frequencies and given the size of the ball, the viscous penetration depth correction to equation~(\ref{eq:Stokes}) is negligible. 

In the ballistic regime, phonons and rotons are responsible for the dissipation, and the decay time contributions at a temperature $T$ are~\cite{Jager_Schuderer_Schoepe_1995b}:

\begin{equation}
    \tau_{\mathrm{ph}} = \frac{45 M \hbar^3 c^4 } {\pi^2 (k_BT)^4 \pi r^2},
\end{equation}

\begin{equation}
    \tau_{\mathrm{rot}} = \frac{6\pi^2 M}{\hbar k_0^4 \exp(-\Delta/k_BT)\pi r^2};
\end{equation}
with $c$ the speed of sound at saturated vapour pressure, $k_0$ the roton-minimum wave number, and $\Delta$ the roton gap. In our case, the roton contribution is mostly negligible. The cross-over between the hydrodynamic and ballistic regimes does not have a good theoretical description, as far as we are aware. We hope that our measurements will stimulate an effort to describe this cross-over. 

The steep $\tau(T)\propto T^{-4}$ dependence can, in fact, be used for precise thermometry in the liquid in the ballistic regime. However, $^3$He impurities present in $^4$He, become the dominating dissipation factor at low enough temperatures.

Dilute $^3$He impurities at low temperatures behave as an ideal gas of  quasiparticle excitations with a thermal velocity
\begin{equation}
    v_{\mathrm{th}} =\sqrt{\frac{2k_B T}{m_3^*}}, 
\end{equation}
and an effective mass $m_3^*=2.64 m_3$, where $m_3$ is the mass of a $^3$He atom~\cite{bardeen1966interactions}.
This allows us to infer the concentration of impurities in $^4$He by calculating the drag on a sphere in Knudsen regime in the limit of slow motion of the sphere~\cite{sengers2014kinetic}. Knudsen number at $T=100$\,mK and at a concentration of $x_3=10^{-8}$ is over 100, so we used the large Knudsen number approximation here, giving for the decay time in this regime:
\begin{equation}
    \tau_{\mathrm{imp}} = \frac{4M}{4.1906\, \pi r^2 n_3 \, m_3^*  v_\mathrm{th} }. 
\end{equation}
\\
We fit the observed $\tau(T)$ dependence with the number density $n_3$ as a free parameter and obtain the concentration of $^3$He impurities $x_3=n_3/n_4\approx4.2\times10^{-8}$ in the purified sample. We have proceeded to contaminating the sample with a known amount of $^3$He,  $x_3'=10^{-7}$. The ring-down time has decreased in full agreement with the amount added as Fig.~\ref{fig:helium_decay} shows.

We note that the cell in our experiments was underfilled, with a substantial area of the free surface. $^3$He in dilute mixtures is known to form a layer of Andreev states on free surfaces with a thickness of $\sim10$\,\AA, thus somewhat depleting the bulk $^3$He concentration~\cite{andreev1966surface, kirichek2024density}. In order to probe the true concentration of $^3$He impurities using the mechanical oscillator immersed in the superfluid, one would need to minimize the free surface, e.g. filling the cell above the filling capillary. A similar probe with a better position resolution might be able to be sensitive to these surface Andreev states and help quantify this phenomenon as well.

\end{document}